\documentclass[a4paper]{article}

\usepackage[english]{babel}
\usepackage[utf8x]{inputenc}
\usepackage[T1]{fontenc}
\usepackage{ upgreek }
\usepackage{gensymb} 
\usepackage{threeparttable}
\usepackage{tabularx}
\usepackage{cite}
\usepackage{mathrsfs}
\usepackage{amssymb}
\usepackage{pifont}
\usepackage{array, caption}
\usepackage{booktabs}
\newtheorem{theorem}{Theorem}{}
\makeatletter
\newcommand{\ssymbol}[1]{^{\@fnsymbol{#1}}}
\makeatother

\usepackage{amsmath}
\usepackage{graphicx}
\usepackage[colorinlistoftodos]{todonotes}
\usepackage[colorlinks=true, allcolors=blue]{hyperref}

\usepackage[a4paper,top=3cm,bottom=2cm,left=3cm,right=3cm,marginparwidth=1.75cm]{geometry}

\usepackage{authblk}
\title{Scalable Mediated Semi-quantum Key Distribution}
\author[1,3]{Chong-Qiang Ye }
\author[2,3]{Jian Li\thanks{Corredponding author:  lijian@bupt.edu.cn}}
\author[2]{Xiu-Bo Chen}
\author[1]{Yan-Yan Hou}
\affil[1]{ School of Artificial Intelligence, Beijing University of Posts Telecommunications, Beijing 100876, China.}
\affil[2]{Information Security Center, State Key Laboratory of Networking and Switching Technology,
Beijing University of Posts and Telecommunications, Beijing 100876, China}
\affil[3]{ Key Laboratory of Cryptography of Zhejiang Province, Hangzhou Normal University, Hangzhou 311121, China}
\begin{document}
  \maketitle

\captionsetup[figure]{labelfont={bf},labelformat={default},labelsep=period,name={Fig.}}
\captionsetup[table]{labelfont={bf},labelformat={default},labelsep=period,name={Table}}

\begin{abstract}
Mediated semi-quantum key distribution (M-SQKD) permits two limited ``semi-quantum'' or ``classical'' users to establish a secret key with the help of a third party (TP), in which TP has fully quantum power and may be untrusted. Several protocols have been studied recently for two-party scenarios, but no one has considered M-SQKD for multi-party scenarios. In this paper, we design a circular M-SQKD protocol based on Bell states, which offers an approach to realizing multiple ``classical'' users' key distribution. Then, we prove the protocol is unconditional security in the asymptotic scenario. The protocol's key rate and noise tolerance can be derived by utilizing the parameters observed in the channel. The results show that our protocol may hold similar security to a fully quantum one. We also compare the proposed protocol with other similar protocols in terms of noise tolerance, qubit efficiency, communication cost and scalability. Finally, the security proof method of this paper may contribute to studying the security of other circular semi-quantum cryptography protocols.
\\
\\
\\
\textbf{Keywords:} Quantum cryptography, Semi-quantum key distribution, Security proof, Scalability

\end{abstract}

\section{Introduction}
\label{intro}
Quantum key distribution (QKD) \cite{1} enables both parties to generate a shared key according to quantum mechanics laws, securely countering the computationally unbounded adversary. Since the first QKD protocol was investigated by Bennett and Brassard, theoretical and experimental studies on QKD have increased significantly \cite{2,3}. Nevertheless, both parties, Alice and Bob, are equipped with complete quantum resources, leading to high overhead for the quantum devices.

Semi-quantum key distribution (SQKD) was first investigated by Boyer et al. \cite{4}, which cut down the quantum capability of one party compared with QKD. In more detail, the SQKD has two parties, one of whom is fully quantum capable of doing anything quantum mechanics allows, while the other is limited to doing the below operations:  (1) \emph{Measure and resend}. Measuring the qubit in the basis $\{|0\rangle,|1\rangle \}$(i.e., $Z$ basis) and then resending an identical qubit to the next recipient; (2) \emph{Reflect}. Reflecting the qubit directly (i.e., learning nothing about its state); (3) \emph{Permute}. Reorder the qubits without disturbing them. The limited party is usually called  ``classical'' or ``semi-quantum''. Obviously, the classical party can only work on the $Z$ basis, which may alleviate the burden of quantum resources. Since then, SQKD has attracted the extensive attention of researchers, and many SQKD protocols and security analysis methods have been proposed \cite{5,6,7,8,9,10,11,12,13}. For example, in 2015, Krawec \cite{6} demonstrated the unconditional security of SQKD protocol and derived the protocol's key rate. After that, many researchers began to study the security of different SQKD protocols. Meanwhile, other types of semi-quantum cryptography protocols were also investigated, such as semi-quantum secure communication \cite{14,15,16}, semi-quantum secret sharing \cite{17,18,19}, and semi-quantum private comparison \cite{20,21,22}. The reader is referred to \cite{23} for a survey, which gave a detailed review of semi-quantum cryptography.

Mediated semi-quantum key distribution (M-SQKD) was first investigated in 2015 by Krawec\cite{24}, which is an essential research focus of SQKD. The primary different between SQKD and M-SQKD is that M-SQKD enables two ``classical'' users (Alice and Bob) to generate a secure key with the help of a third party (TP), while SQKD allows the keys to be shared between quantum and ``classical'' users. M-SQKD can be regarded as a basic protocol in the field of semi-quantum cryptography. So far, several M-SQKD protocols have been proposed with various quantum states and transport structures \cite{25,26,27,28,29,30,31,32}. In Ref. \cite{25}, Krawec improved the key rate of the original M-SQKD protocol by using an alternative method of proof. After that, Liu et al. \cite{26} presented a new M-SQKD protocol based on entanglement swapping, where the ``classical'' users do not require quantum measurement capability. Lin et al. \cite{27} presented an M-SQKD protocol with single photons instead of entangled ones to reduce the necessary quantum resources. Recently, Chen et al. \cite{28} designed an M-SQKD protocol using single-particle states. Unlike the previous protocols, this protocol adopts the mode of circular transmission (i.e., the qubits travel from TP to Alice, to Bob, then back to TP). In 2019, Krawec proposed a multi-party M-SQKD protocol where two (or more) adversarial quantum servers are used to help two ``classical'' users establish a secret key \cite{29}. In 2022, Guskind et al. \cite{30} improved the M-SQKD protocol's efficiency without requiring additional users' additional capabilities. In the above mentioned, only a few M-SQKD protocols provided security proof \cite{24,25,30}, while others only analyzed that protocols can resist certain types of attacks. In particular, no one has yet to consider how to implement key distribution for multiple "classical" users.

In this paper, we attempt to design an M-SQKD protocol that works for multiple ``classical'' users to establish the secret keys. Due to the scalability of circular transmission, this provides a way to implement key distribution for multiple classical users. Naturally, we propose a circular M-SQKD protocol based on Bell states, which is suitable for multi-party scenarios. To prove security, we figure out a lower bound of the protocol's key rate and noise tolerance (i.e., error tolerance), by utilizing the specific parameters Alice and Bob may estimate. Indeed, if TP is semi-honest (i.e., TP fully complies with the requirements of the protocol, but otherwise, she can do anything she likes), using the key rate bound in this paper, Alice and Bob can obtain a secret key when the noise is smaller than 16.02\%. If TP is untrusted (TP can do anything she wants, including preparing fake particles and publishing fake results), they can obtain a secret key when the noise is less than 13.19\%. The results show that our protocol's noise tolerance is close to previous protocols \cite{24,25,30}. Even in the worst case (i.e., TP is untrusted), the noise tolerance is also better than previous M-SQKD protocols. Furthermore, we evaluate the performance of the protocol in terms of noise tolerance, qubit efficiency, communication cost and scalability. Compared with similar protocols, our protocol has comparable advantages in the above four aspects.

In summary, the contributions of this work are as follows. First, we propose a circular M-SQKD protocol with good scalability, which offers an approach to realizing multiple ``classical'' users' key distribution. Second, we give a detailed security proof of this circular M-SQKD protocol. The results show that the proposed protocol has similar security as the full quantum protocol. 




\section{ The Proposed M-SQKD Protocol}
Let's use two-party scenarios as an example for illustration (the multi-party scenarios will be discuss later). The ``classical'' users, Alice and Bob, are only allowed to reflect the qubits to the next recipient; or measure and resend the qubits to the next recipient on $Z$ basis. They desire to generate a secure key with a quantum TP's help, who may be untrusted. The detailed process is given in the following.

\textbf{Step 1:} TP prepares $8(N+\delta)$ Bell states where each state is $|\phi^+\rangle=\frac{1}{\sqrt2}(|00\rangle+|11\rangle)$ and $\delta$ is a fixed parameter. Subsequently, she sends the second qubit of each Bell state to Alice.	
		 			
\textbf {Step 2:} Upon receiving the qubit, Alice randomly decides to reflect it to Bob or measure and resend it to Bob.

\textbf {Step 3:} Similarly, Bob randomly decides to reflect or to measure and resend the received qubit to TP.

\textbf {Step 4:} TP uses the Bell basis to measure the received qubit and the corresponding qubit in her hand. Then, TP announces her measurements to Alice and Bob.

\textbf {Step 5:} After TP releases her measurements, Alice and Bob divulge their choices through the authenticated classical channels. According to their choices, the following three cases are available. 
\begin{itemize}
\item \textbf{Case 1}: When Alice and Bob reflect the received qubit, TP should always publish $|\phi^+\rangle$. This case is used for checking the honesty of TP and eavesdropping. If error rate surpasses the threshold, the protocol ends.

\item \textbf{Case 2}: When they measure and resend the received qubit, TP will publish the result as either $|\phi^+\rangle$ or $|\phi^-\rangle$. If the result of TP is not $|\phi^+\rangle$ or $|\phi^-\rangle$, they will discard this case.


\item \textbf{Case 3}: When they perform different operations on the received qubit, this case will be discarded.

\end{itemize}

\textbf {Step 6:} Alice and Bob select a small subset of Case 2 and divulge their measurements for eavesdropping detection. Once the error rate is over the threshold, the protocol terminates. If not, the protocol continues.

\textbf {Step 7:} After the classical post-processing (e.g., privacy amplification), two ``classical'' users can obtain the final secure key.

\section{Security}

In this section, we will give a detailed proof of security. Here we only consider the threat posed by an adversarial TP instead of outside eavesdroppers because outside eavesdroppers' attacks can be absorbed into the TP's attacks. Besides, according to \cite{33} and \cite{34}, security against collective attacks is enough to demonstrate security against arbitrary general attacks. Thereby, we focus on proving security against collective attacks.
 
In our protocol, the raw keys are only established in the event that Alice and Bob choose to measure, and that TP announces $|\phi^-\rangle$. Under collective attacks, the system of Alice, Bob, and TP can be described as  
\begin{equation}
\rho_{ABT}=\sum_{i,j}|i,j\rangle \langle i,j|_{AB}\otimes\rho_{T},
\end{equation} 
where $\rho_{T}$ denotes the state of TP's ancilla. Let $N$ denote the size of Alice and Bob's raw key, $\ell(n)$ denote the length of secure key. Based on \cite{2}, our protocol's key rate in the asymptotic scenario is
\begin{equation}
r =\lim_{N\to \infty} \frac{\ell(N)}{N} = {\rm inf} \lbrack S(A|T)-H(A|B) \rbrack,
\end{equation} 
where $S(\cdot)$ denotes the Von Neumann entropy, and $H(\cdot)$ means the Shannon entropy.  Observing the system's parameters, the conditional entropy of $S(A|T)$ and $H(A|B)$ can be derived. If $r>0$, the protocol can obtain a secure secret key. Hence, the goal of the next stage is to analyze TP's attack strategy and compute the key rate $r$.

\subsection{The scenario of semi-honest TP}
We first consider a particular scenario where TP is semi-honest. In this scenario, TP follows the protocol description thoroughly, but otherwise, she can do anything she likes. As for channel noise, we assume that its effects can be incorporated into the errors caused by TP attacks\cite{6}.

\subsubsection{The attack strategy of semi-honest TP }
To catch Alice and Bob's raw key, the attack strategy of TP can be modeled as follows (Also see Fig. 1). Here, we first consider Alice and Bob both measure and resend the received qubit.

\begin{figure}[htb]
\centering\includegraphics[width=4.2in]{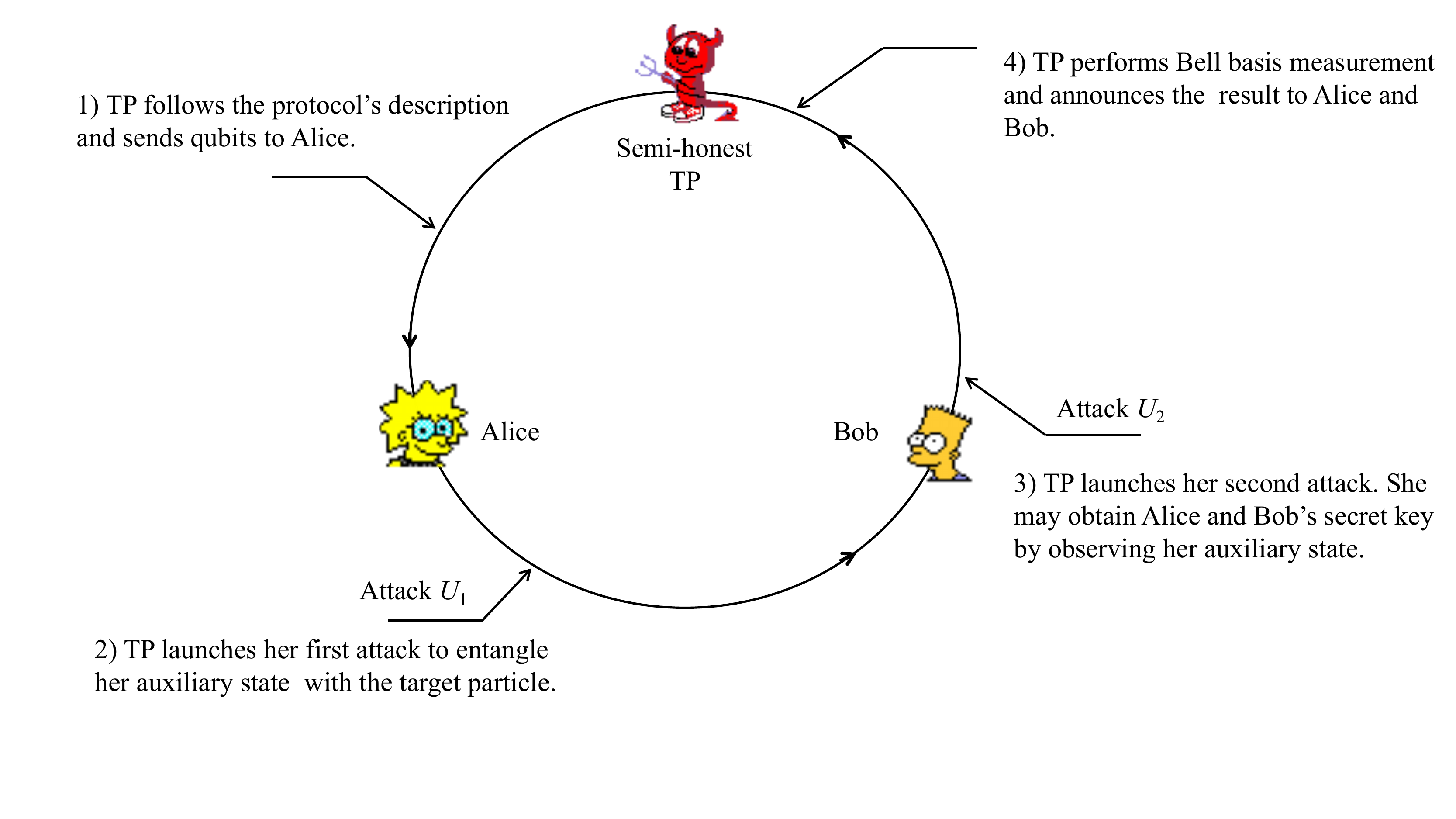}
\caption{A diagram of semi-honest TP's attack strategy. TP follows the protocol description thoroughly, but otherwise, she can do anything she likes. She will perform $U_1$ and $U_2$ attacks on the qubits of Alice and Bob, respectively, and may obtain the raw key by observing her auxiliary particles. }
\label{fig:1}       
\end{figure}

1)  According to the protocol's description, TP first sends the second qubit of $|\phi^+\rangle$ to Alice. After that, Alice measures and resends the received qubit. Thus, the density operator describing the system of Alice and TP is
\begin{equation}
\begin{aligned}
\rho_{1}=\frac{1}{2}|00\rangle \langle 00|_{HA}+\frac{1}{2}|11\rangle \langle 11|_{HA}
\end{aligned}
\end{equation} 
where $H$ denotes the qubit left in TP's hands, and $A$ denotes Alice's measurement result.  

2) TP launches her first attack on the qubits from Alice. Specifically, she performs the $U_1$ operation on the Alice's qubit and her auxiliary state $|0\rangle_{T}$. The effect of $U_1$ is
\begin{eqnarray}
& U_1\left|0,0\right\rangle=|0,E_0\rangle+|1,E_1\rangle,\\ 
& U_1\left|1,0\right\rangle=|1,E_3\rangle+|0,E_2\rangle.
\end{eqnarray}
After TP's attack, the system becomes
\begin{equation}
\begin{aligned}
\rho_{2}&=\frac{1}{2}|0\rangle \langle 0|_{H}\otimes(|0,E_{0}\rangle \langle 0,E_{0}|_{A^\prime T} +|1,E_{1}\rangle \langle 1,E_{1}|_{A^\prime T} )\\
&+\frac{1}{2}|1\rangle \langle 1|_{H}\otimes(|0,E_{2}\rangle \langle 0,E_{2}|_{A^\prime T} +|1,E_{3}\rangle \langle 1,E_{3}|_{A^\prime T} ),
\end{aligned}
\end{equation} 
Here, $|E_0\rangle$,$|E_1\rangle$,$|E_2\rangle$, and $|E_3\rangle$ are pure states determined by $U_1$. $A^\prime$ means the qubit of Alice after TP's attack and it will be received by Bob.

3) Bob measures and resends the received qubit. The system transforms to
\begin{equation}
\begin{aligned}
\rho_{3}&=\frac{1}{2}|00\rangle \langle 00|_{HA}\otimes(|0,E_{0}\rangle \langle 0,E_{0}|_{BT} +|1,E_{1}\rangle \langle 1,E_{1}|_{B T} )\\
&+\frac{1}{2}|11\rangle \langle 11|_{HA}\otimes(|0,E_{2}\rangle \langle 0,E_{2}|_{B T} +|1,E_{3}\rangle \langle 1,E_{3}|_{BT} ),
\end{aligned}
\end{equation} 
where $B$ denotes Bob's measurement result. Then, TP launches her second attack. Specifically, she performs the $U_2$ operation on the Bob's qubit and her auxiliary state $|E_k\rangle$. The effect of $U_2$ is as follows: 
\begin{equation}
\begin{aligned}
U_2|i,E_k\rangle=\sum_{j=0}^{1}|j,E^{j}_{i,k}\rangle, 
\end{aligned}
\end{equation}
where $i,j\in\{0,1\}$ and $k\in\{0,1,2,3\}$. Thus, the system of Alice, Bob, and TP becomes
\begin{equation}
\begin{aligned}
\rho_4&=\frac{1}{2}\big(|0000\rangle\langle 0000|_{HABB^\prime}\otimes|E^0_{0,0}\rangle \langle E^0_{0,0}|_T\\
&\quad\quad +|0001\rangle\langle 0001|_{HABB^\prime}\otimes|E^1_{0,0}\rangle \langle E^1_{0,0}|_T\\
&\quad\quad +|0010\rangle\langle 0010|_{HABB^\prime}\otimes|E^0_{1,1}\rangle \langle E^0_{1,1}|_T\\
&\quad\quad +|0011\rangle\langle 0011|_{HABB^\prime}\otimes|E^1_{1,1}\rangle \langle E^1_{1,1}|_T\\
&\quad\quad +|1100\rangle\langle 1100|_{HABB^\prime}\otimes|E^0_{0,2}\rangle \langle E^0_{0,2}|_T\\
&\quad\quad +|1101\rangle\langle 1101|_{HABB^\prime}\otimes|E^1_{0,2}\rangle \langle E^1_{0,2}|_T\\
&\quad\quad +|1110\rangle\langle 1110|_{HABB^\prime}\otimes|E^0_{1,3}\rangle \langle E^0_{1,3}|_T\\
&\quad\quad +|1111\rangle\langle 1111|_{HABB^\prime}\otimes|E^1_{1,3}\rangle \langle E^1_{1,3}|_T
\big),
\end{aligned}
\end{equation}
where $B^\prime$ means the qubit received by TP.

4) TP uses the Bell basis to measure the received qubit $B^\prime$ and qubit $H$ in her hand, and publishes the outcomes to Alice and Bob. Subsequently, according to the statement of TP and the qubits observed by Alice and Bob, it will get:
\begin{equation}
\begin{aligned}
&p^{c}_{0,0}=\langle E^0_{0,0}| E^0_{0,0}\rangle,&  
&p^{w}_{0,0}=\langle E^1_{0,0}| E^1_{0,0}\rangle,&\\
&p^{c}_{0,1}=\langle E^{0}_{1,1}| E^{0}_{1,1}\rangle,& 
&p^{w}_{0,1}=\langle E^{1}_{1,1}| E^{1}_{1,1}\rangle, &\\
&p^{c}_{1,0}=\langle E^1_{0,2}| E^1_{0,2}\rangle,&  
&p^{w}_{1,0}=\langle E^0_{0,2}| E^0_{0,2}\rangle,&\\
&p^{c}_{1,1}=\langle E^{1}_{1,3}| E^{1}_{1,3}\rangle,& 
&p^{w}_{1,1}=\langle E^{0}_{1,3}| E^{0}_{1,3}\rangle,
\end{aligned}
\end{equation}
where $p^{c}_{i,j}$ means the probability that the qubits $B^\prime$ and $H$ are consistent (i.e., TP's statement is $|\phi^+\rangle$ or $|\phi^-\rangle$), and the qubits observed by Alice and Bob are $|i\rangle$ and $|j\rangle$, while $p^{w}_{i,j}$ represents the probability that the qubits $B^\prime$ and $H$ are inconsistent, and the qubits observed by Alice and Bob are $|i\rangle$ and $|j\rangle$.

However, Alice and Bob only accept the event where TP announces $|\phi^+\rangle$ or $|\phi^-\rangle$. Conditioning on this event and tracing out $B^\prime$ and $H$, the final contributing raw key system is
\begin{equation}
\begin{aligned}
\rho_{ABT}&=\frac{1}{N}\big(|00\rangle\langle 00|_{AB}\otimes|E^0_{0,0}\rangle \langle E^0_{0,0}|_T\\
&\quad\quad +|01\rangle\langle 01|_{AB}\otimes|E^0_{1,1}\rangle \langle E^0_{1,1}|_T\\
&\quad\quad +|10\rangle\langle 10|_{AB}\otimes|E^1_{0,2}\rangle \langle E^1_{0,2}|_T\\
&\quad\quad +|11\rangle\langle 11|_{AB}\otimes|E^1_{1,3}\rangle \langle E^1_{1,3}|_T
\big),
\end{aligned}
\end{equation}
where $N$ is the normalization term: $N=\langle E^0_{0,0}| E^0_{0,0}\rangle+\langle E^0_{1,1}| E^0_{1,1}\rangle+\langle E^1_{0,2}| E^1_{0,2}\rangle+\langle E^1_{1,3}| E^1_{1,3}\rangle $. Let $p_{i,j}$ denote the probability that Alice and Bob's raw keys are $i$ and $j$.
From (11), we have
\begin{equation}
\begin{aligned}
&p_{0,0}=\frac{1}{N}\langle E^{0}_{0,0}| E^{0}_{0,0}\rangle,&  
&p_{0,1}=\frac{1}{N}\langle E^{0}_{1,1}| E^{0}_{1,1}\rangle,&\\
&p_{1,0}=\frac{1}{N}\langle E^{1}_{0,2}| E^{1}_{0,2}\rangle,& 
&p_{1,1}=\frac{1}{N}\langle E^{1}_{1,3}| E^{1}_{1,3}\rangle, 
\end{aligned}
\end{equation}
TP may obtain the raw keys by observing the states of her auxiliary state. Recall the definition of the key rate, as long as $r=S(A|T)-H(A|B)> 0$, a secure secret key can be distilled. Thus, in the next step, we have to calculate the range of $r$.

\subsubsection{Key rate derivation}

Obviously, the calculation of $H(A|B)$ is simple. According to the (12), $H(A|B)$ is calculated:
\begin{equation}
\begin{aligned}
H(A|B)&=H(A,B)-H(B)\\
&=H(p_{0,0},p_{0,1},p_{1,0},p_{1,1})-H(p_{0,0}+p_{1,0},p_{1,1}+p_{0,1}).
\end{aligned}
\end{equation}

To bound $S(A|T)$, we introduce a new auxiliary system $F$ to simplify the calculation process. According to the strong sub additivity, we have $S(A|T)\ge S(A|TF)$.
Thereby, we can obtain a lower bound of $r$
\begin{equation}
r=S(A|T)-H(A|B)\ge S(A|TF)-H(A|B).
\end{equation}
The new system $F$ is spanned by $\{ |C\rangle,|W\rangle \}$, where $|C\rangle$ means the raw key of Alice and Bob is consistent, while $|W\rangle$ represents their raw key is not consistent. Then the new mixed system is:
\begin{equation}
\begin{aligned}
\rho_{ABTF}&=\frac{1}{N}\big(|00\rangle\langle 00|_{AB}\otimes|E^0_{0,0}\rangle \langle E^0_{0,0}|_T \otimes |C\rangle \langle C|_F\\
&\quad\quad +|01\rangle\langle 01|_{AB}\otimes|E^0_{1,1}\rangle \langle E^0_{1,1}|_T\otimes |W\rangle \langle W|_F\\
&\quad\quad +|10\rangle\langle 10|_{AB}\otimes|E^1_{0,2}\rangle \langle E^1_{0,2}|_T\otimes |W\rangle \langle W|_F\\
&\quad\quad +|11\rangle\langle 11|_{AB}\otimes|E^1_{1,3}\rangle \langle E^1_{1,3}|_T \otimes |C\rangle \langle C|_F
\big),
\end{aligned}
\end{equation}
Tracing out $B$ from $\rho_{ABTF}$, $\rho_{ATF}$ can be written as a diagonal matrix, where the elements are $p_{i,j}$. Thereby:
\begin{equation}
S(ATF)=S(\rho_{ATF})= H(p_{0,0}, p_{0,1}, p_{1,0},p_{1,1}).
\end{equation}
Next, we calculate $S(TF)$. Tracing out $A$ and $B$ from $\rho_{ABTF}$ yields:
\begin{equation}
\begin{aligned}
\rho_{TF} &= \frac{1}{N}(|E^0_{0,0}\rangle \langle E^0_{0,0}|_T +|E^1_{1,3}\rangle \langle E^1_{1,3}|_T)\otimes |C\rangle \langle C|_F\\
&+\frac{1}{N}(|E^0_{1,1}\rangle \langle E^0_{1,1}|_T +|E^1_{0,2}\rangle \langle E^1_{0,2}|_T )\otimes |W\rangle \langle W|_F.
\end{aligned}
\end{equation}
Let $\xi_1=\frac{1}{N}\big(\langle E^0_{0,0}|E^0_{0,0}\rangle+\langle E^1_{1,3}|E^1_{1,3}\rangle \big)=p_{0,0}+p_{1,1}$, $\xi_2=\frac{1}{N}\big(\langle E^0_{1,1}|E^0_{1,1}\rangle+\langle E^1_{0,2}|E^1_{0,2}\rangle \big)=p_{0,1}+p_{1,0}$. Then (17) can be rewritten as
\begin{equation}
\rho_{TF} =\xi_1 \sigma_1 \otimes |C\rangle \langle C|_F +\xi_2 \sigma_2 \otimes |W\rangle \langle W|_F,
\end{equation}
where
\begin{equation}
\begin{aligned}
&\sigma_1 = \frac{|E^0_{0,0} \rangle \langle E^0_{0,0}|+| E^{1}_{1,3}\rangle \langle E^{1}_{1,3}|}{\langle E^0_{0,0}|E^0_{0,0}\rangle+\langle E^1_{1,3}|E^1_{1,3}\rangle}, \quad
 \sigma_2= \frac{|E^0_{1,1}\rangle \langle E^0_{1,1}|+|e^1_{0,2}\rangle \langle E^1_{0,2}|}{\langle E^0_{1,1}|e^0_{1,1}\rangle+\langle E^1_{0,2}|E^1_{0,2}\rangle}.
\end{aligned}
\end{equation}
According to the joint entropy theorem of Von Neumann entropy (For detailed proof, see \cite{6}), $S(\rho_{TF})$ can be derived as 
\begin{equation}
\begin{aligned}
S(TF)=S(\rho_{TF})= H(p_{0,0}+p_{1,1}, p_{1,0}+p_{0,1})+ (p_{0,0}+p_{1,1}) S(\sigma_1) + (p_{1,0}+p_{0,1}) S(\sigma_2)
\end{aligned}.
\end{equation}
Since $\sigma_1$ and $\sigma_2$ are two-dimensional systems, we have $S(\sigma_1)\le 1$ and $S(\sigma_2)\le 1$.  
Thus, the upper bound $S(TF)$ can be calculated as follow:
\begin{equation}
\begin{aligned}
S(TF) &\le H(p_{0,0}+p_{1,1}, p_{1,0}+p_{0,1})+ (p_{0,0}+p_{1,1}) S(\sigma_1) +p_{1,0}+p_{0,1}.
\end{aligned}
\end{equation}
 Let $|E^0_{0,0}\rangle=\alpha|h\rangle$ and  
 $|E^{1}_{1,3}\rangle=\beta\left|h\right\rangle + \theta \left|\zeta\right\rangle$, where $\alpha,\beta,\theta \in \mathbb{C}$, $\left \langle h|h\right\rangle=\left \langle\zeta|\zeta\right\rangle =1 $ and $\left\langle h|\zeta\right\rangle =0$. Then, we have $\langle E^{0}_{0,0}|E^0_{0,0}\rangle=\|\alpha\|^2$, $\langle E^{1}_{1,3}|E^1_{1,3}\rangle=\|\beta\|^2+\|\theta\|^2$, and $\langle E^{0}_{0,0}|E^1_{1,3}\rangle=\alpha^*\beta$. Thus, $\sigma_1$ can be described as 
\begin{equation}
\sigma_1= 
 \frac{1}{N\cdot\xi_1}
 \begin{pmatrix}
\|\alpha\|^2+\|\beta\|^2  \quad  & \beta\theta^* \\
\beta^*\theta \quad  & \|\theta\|^2
\end{pmatrix}.
\end{equation}
After that the eigenvalues of $\sigma_1 $ can be calculated:
\begin{equation}
	\lambda_{\pm}= \frac{1}{2} \pm \frac{1}{2N\cdot \xi_1}\sqrt{N^2(p_{0,0}-p_{1,1})^2+4 \langle E^0_{0,0}|E^{1}_{1,3}\rangle^2}.
\end{equation}
Thus, $S(\sigma_1)=-\lambda_{+}log\lambda_{+}-\lambda_{-}log\lambda_{-}=H(\lambda_{+},\lambda_{-})$. Using (14), (16), and (21), the key rate is simplified as
\begin{equation}
\begin{aligned}
	r &\ge S(A|TF)-H(A|B) \ge H(p_{0,0}+p_{1,0},p_{0,1}+p_{1,1})-p_{1,0}-p_{0,1}\\
	&- (p_{0,0}+p_{1,1}) S(\sigma_1)-H(p_{0,0}+p_{1,1}, p_{1,0}+p_{0,1}). 
\end{aligned}
\end{equation}
What remains is to bound $S(\sigma_1)$. It's not hard to see that $\lambda_{+}\ge \frac{1}{2}$, and that as $\langle E^0_{0,0}|E^{1}_{1,3}\rangle^2$ decreases $S(\sigma_1)$ will increase. Therefore, to obtain a lower bound of $r$, we should identify a lower bound on $\langle E^0_{0,0}|E^{1}_{1,3}\rangle^2$.

Observing the case that Alice and Bob both reflect the received qubit, we may bound $\langle E^0_{0,0}|E^{1}_{1,3}\rangle^2$. Specifically, after TP's attacks $U_1$ and $U_2$, the system will be
\begin{equation}
\begin{aligned}
U_2U_1|\phi^+\rangle|0\rangle_T
&=\frac{1}{2}\big[|\phi^+\rangle(|F_0\rangle+|F_3\rangle)+|\phi^-\rangle(|F_0\rangle-|F_3\rangle)\\
&\quad+|\psi^-\rangle(|F_1\rangle-|F_2\rangle)+|\psi^+\rangle(|F_1\rangle+|F_2\rangle)\big],
\end{aligned}
\end{equation}
where 
\begin{equation}
\begin{aligned}
&|F_0\rangle=|E^0_{0,0}\rangle+|E^0_{1,1}\rangle, \quad |F_1\rangle=|E^1_{0,0}\rangle+|E^1_{1,1}\rangle,\\
&|F_2\rangle=|E^0_{0,2}\rangle+|E^0_{1,3}\rangle, \quad |F_3\rangle=|E^1_{0,2}\rangle+|E^1_{1,3}\rangle.
\end{aligned}
\end{equation}
Then, TP measures the qubits with Bell basis and announces her measurements. If the measurement is not $|\phi^+\rangle$, Alice and Bob know there is an error raised. Here, we use $p_{\psi^-}$, $p_{\psi^+}$, and $p_{\phi^-}$ to represent the probabilities that TP's outcomes are $|\psi^-\rangle$, $|\psi^+\rangle$, and $|\phi^-\rangle$, respectively. According to (25),  it's easy to get
\begin{equation}
\begin{aligned}
&p_{\psi^-}=\frac{1}{4}\langle F_1|F_1\rangle -\frac{1}{2}Re\langle F_1|F_2\rangle + \frac{1}{4}\langle F_2|F_2\rangle,\\
&p_{\psi^+}=\frac{1}{4}\langle F_1|F_1\rangle +\frac{1}{2}Re\langle F_1|F_2\rangle + \frac{1}{4}\langle F_2|F_2\rangle ,\\
&p_{\phi^-}=\frac{1}{4}\langle F_0|F_0\rangle -\frac{1}{2}Re\langle F_0|F_3\rangle + \frac{1}{4}\langle F_3|F_3\rangle .\\
\end{aligned}
\end{equation}
Note that $U_1$ and $U_2$ are unitary, which mean that $\langle F_0|F_0\rangle+\langle F_1|F_1\rangle=\langle F_2|F_2\rangle+\langle F_3|F_3\rangle=1$. After some algebra, we have
\begin{equation}
 p_{\psi^+}+p_{\psi^-} + p_{\phi^-} =1-\frac{1}{2}Re\langle F_0|F_3\rangle-\frac{1}{4}\langle F_0|F_0\rangle-\frac{1}{4}\langle F_3|F_3\rangle.
\end{equation}
Using (26) into (28), thus
\begin{equation}
\begin{aligned}
Re\langle E^0_{0,0}|E^1_{1,3}\rangle&=2-2(p_{\psi^+}+p_{\phi^-}+p_{\psi^-})\\
&\quad -Re(\langle E^0_{0,0}|E^1_{0,2}\rangle+\langle E^0_{1,1}| E^1_{1,3}\rangle+\langle E^0_{1,1}|E^1_{0,2}\rangle)\\
&\quad-\frac{1}{2}Re(\langle E^0_{0,0}|E^0_{0,0}\rangle+\langle E^0_{1,1}| E^0_{0,0}\rangle+\langle E^0_{0,0}|E^0_{1,1}\rangle+\langle E^0_{1,1}|E^0_{1,1}\rangle)\\
&\quad-\frac{1}{2}Re(\langle E^1_{0,2}|E^1_{0,2}\rangle+\langle E^1_{1,3}| E^1_{0,2}\rangle+\langle E^1_{0,2}|E^1_{1,3}\rangle+\langle E^1_{1,3}|E^1_{1,3}\rangle).
\end{aligned}
\end{equation}
Then, according to the Cauchy-Schwarz inequality (i.e., $|Re\langle a|b\rangle |\le |\langle a|b \rangle |\le \sqrt{\langle a|a\rangle \langle b|b\rangle}$) and (10), the lower bound of $Re\langle E^0_{0,0}|E^1_{1,3}\rangle$ can be obtained
\begin{equation}
\begin{aligned}
Re\langle E^0_{0,0}|E^1_{1,3}\rangle& \ge 2-2(p_{\psi^+}+p_{\phi^-}+p_{\psi^-})\\
&\quad -\big(\sqrt{p^{c}_{0,0}p^{c}_{1,0}}+\sqrt{p^{c}_{0,1}p^{c}_{1,1}}+\sqrt{p^{c}_{0,1}p^{c}_{1,0}} \big)\\
&\quad-\frac{1}{2}\big(\sqrt{p^{c}_{0,0}p^{c}_{0,0}}+\sqrt{p^{c}_{0,1}p^{c}_{0,0}}+\sqrt{p^{c}_{0,0}p^{c}_{0,1}}+\sqrt{p^{c}_{0,1}p^{c}_{0,1}} \big)\\
&\quad-\frac{1}{2}\big(\sqrt{p^{c}_{1,0}p^{c}_{1,0}}+\sqrt{p^{c}_{1,1}p^{c}_{1,0}}+\sqrt{p^{c}_{1,0}p^{c}_{1,1}}+\sqrt{p^{c}_{1,1}p^{c}_{1,1}} \big).
\end{aligned}
\end{equation}

From (23) and (30), we can bound $S(\sigma_1)$. So far, we have derived the bounds of $H(A|B)$, $S(ATF)$, and $S(TF)$ which depend on the probabilities observed in (10), (12) and (27).  Hence, the key rate can be derived. 

\subsubsection{Key rate evaluation }
The system's errors (or noise) are all caused by the attacks from TP. Thus, we need to model the errors introduced by TP attacks. In particular, suppose that the errors introduced by TP's attacks can be characterized as (This is a typical assumption applied to the security analysis):
\begin{itemize}
\item [1.] Let $Q$ express the probability that Alice and Bob's measurement results are different.
\item [2.] Let $Q_M$ represent the probability that TP announces the wrong message when Alice and Bob choose measurement. For example, they observe the same quantum state, but TP publishes $|\psi^+\rangle$ or $|\psi^-\rangle$.
\item [3.] Let $Q_R$ represent the probability that TP announces the wrong result when Alice and Bob reflect the qubits. That is, $p_{\psi^+}+p_{\psi^-}+p_{\phi^-}=Q_R$
\end{itemize}
Note that the error rates for $Q$, $Q_M$, and $Q_R$ can be easily observed in this protocol. Thereby, (10) can be represented by $Q$ and $Q_M$ as follows
\begin{equation}
\begin{aligned}
&p^{c}_{0,0}=p^{c}_{1,1}=(1-Q)(1-Q_M),& 
&p^{c}_{1,0}=p^{c}_{0,1}=QQ_M,&\\
&p^{w}_{0,0}=p^{w}_{1,1}=Q_M(1-Q),&
&p^{w}_{1,0}=p^{w}_{0,1}=(1-Q_M)Q.
\end{aligned}
\end{equation}
Then (12) can be rewritten as 
\begin{equation}
\begin{aligned}
&p_{0,0}=\frac{(1-Q)(1-Q_M)}{2(1-Q)(1-Q_M)+2QQ_M},&  
&p_{0,1}=\frac{QQ_M}{2(1-Q)(1-Q_M)+2QQ_M},&\\
&p_{1,0}=\frac{QQ_M}{2(1-Q)(1-Q_M)+2QQ_M},& 
&p_{1,1}=\frac{(1-Q)(1-Q_M)}{2(1-Q)(1-Q_M)+2QQ_M}.
\end{aligned}
\end{equation}
Moreover, (30) can be expressed as 
\begin{equation}
\begin{aligned}
Re\langle E^0_{0,0}|E^1_{1,3}\rangle& \ge 2-2 Q_R-4\sqrt{(1-Q)(1-Q_M)QQ_M}\\
&\quad -2QQ_M-(1-Q)(1-Q_M),
\end{aligned}
\end{equation}
so that $S(\sigma_1)$ also can be denoted by $Q$, $Q_M$, and $Q_R$.
\begin{figure}[h]
\centering\includegraphics[width=3.5in]{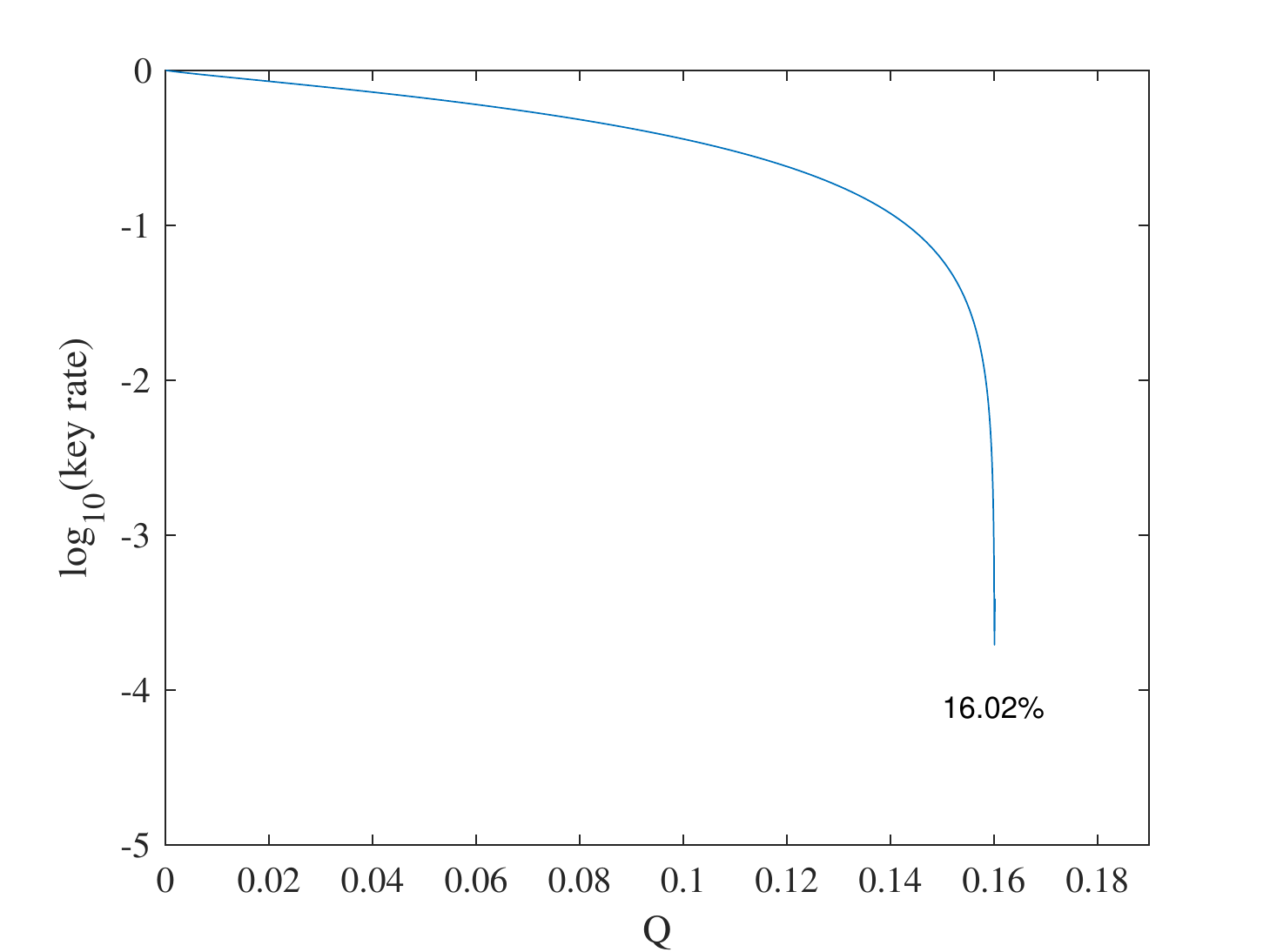}
\caption{Key rate of the proposed protocol when TP is semi-honest. This graph assumes the error rates of $Q_R$, $Q_M$, and $Q$ are equal, i.e., $Q_R=Q_M=Q$. Observe that the protocol can work with the noise of less than 16.02\%.}
\label{fig:2}       
\end{figure}

Thus, the bounds of $H(A|B)$, $S(ATF)$, and $S(TF)$ are all denoted by the error rates $Q$, $Q_M$, and $Q_R$. Putting everything together into the (24), the key rate $r$ can be estimated. In particular, considering the situation when $Q=Q_M=Q_R$, the key rate $r$ only depends on the variable $Q$. The relationship between $r$ and $Q$ is shown in Fig. 2. As long as $Q \le 16.02\% $, we have $r> 0$, that is, a secure key can be obtained by Alice and Bob.


\subsection{The scenario of untrusted TP}
In the above subsection, we discussed a particular scenario where TP is semi-honest. However, in practice, it is often not so ideal. Here, we consider the worst scenario where TP is untrusted. In this scenario, TP can do anything she likes, including preparing fake particles and publishing fake results. 

\subsubsection{The attack strategy of untrusted TP }

TP is adversarial, so she unnecessarily follows the protocol description to prepare Bell state, but instead any arbitrary state $|\phi\rangle=\sum_{i=0}^{1}|i\rangle \otimes|g_i\rangle$, where the $|g_i\rangle$ are TP's private ancilla and the normalized parameters of $|\phi\rangle$ are incorporated into 
 $|g_i\rangle$. 
Similar to Sect. 3.1, we also first consider the case where Alice and Bob both measure and resend the received qubit. To obtain the raw key, TP's attack strategy is modeled as follows (also see Fig. 3).

\begin{figure}[ht]
\centering\includegraphics[width=4.5in]{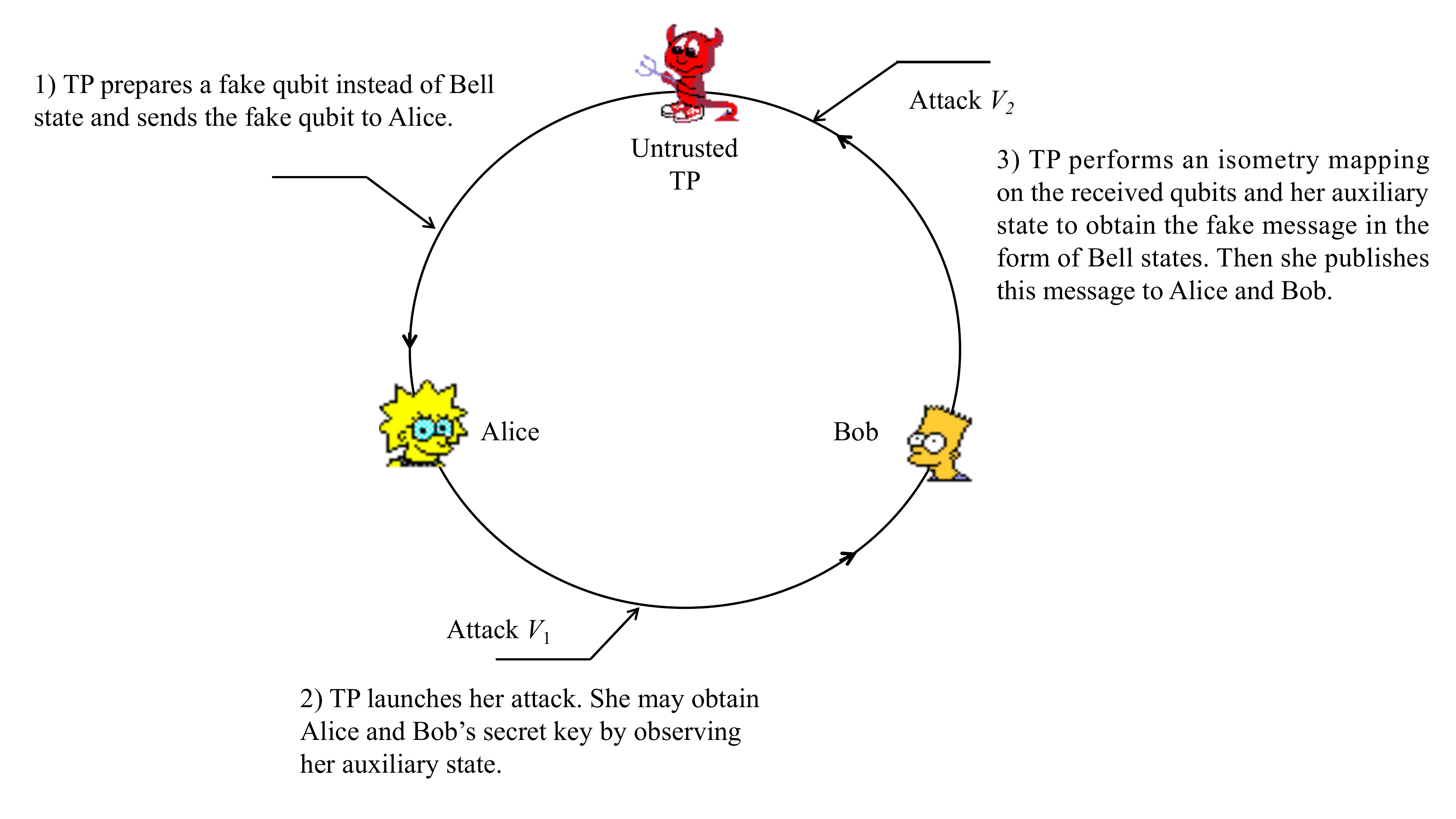}
\caption{A diagram of untrusted TP's attack strategy. TP can do anything she likes, including preparing fake particles and publishing fake results. TP may obtain the raw key by observing her auxiliary particles. }
\label{fig:4}       
\end{figure}

1) TP sends the qubit $|i\rangle$ to Alice. Then, Alice measures and resends the received qubit. Thus the density operator describing the system can be expressed as
\begin{equation}
\begin{aligned}
\rho^{\prime}_1=|0\rangle \langle 0| \otimes |g_0\rangle \langle g_0|+ |1\rangle \langle 1| \otimes |g_1\rangle \langle g_1|,
\end{aligned}
\end{equation}

2) Then, TP executes the $V_1$ operation on the qubit from Alice and her auxiliary state, where $V_1$ is defined as $V_1|i,g_i\rangle=\sum_{j=0}^{1}|j,g_{i,j}\rangle$. Thus the system of Alice, transmitted qubit, and TP's auxiliary qubit is
\begin{equation}
\begin{aligned}
\rho^{\prime}_2 &=|00\rangle \langle 00|_{AA^\prime}\otimes|g_{0,0}\rangle \langle g_{0,0}|_{T} + |01\rangle \langle 01|_{AA^\prime}\otimes|g_{0,1}\rangle \langle g_{0,1}|_{T}\\
& +|10\rangle \langle 10|_{AA^\prime}\otimes|g_{1,0}\rangle \langle g_{1,0}|_{T} + |11\rangle \langle 11|_{AA^\prime}\otimes|g_{1,1}\rangle \langle g_{1,1}|_{T},
\end{aligned}
\end{equation}
where $A$ denotes Alice's measurement result, $A^\prime$ represents the state received by Bob, and $T$ means TP's auxiliary state. After that, Bob measures and resends the received qubit. Then, the system of Alice, Bob and TP is 
\begin{equation}
\begin{aligned}
\rho^{\prime}_3&=|000\rangle \langle 000|_{ABB^\prime}\otimes|g_{0,0}\rangle \langle g_{0,0}|_{T} \\
& +|011\rangle \langle 011|_{ABB^\prime}\otimes|g_{0,1}\rangle \langle g_{0,1}|_{T}\\
&+|100\rangle \langle 100|_{ABB^\prime}\otimes|g_{1,0}\rangle \langle g_{1,0}|_{T}\\
&+|111\rangle \langle 111|_{ABB^\prime}\otimes|g_{1,1}\rangle \langle g_{1,1}|_{T},
\end{aligned}
\end{equation}
where $B$ denotes Bob's measurement result, and $B^\prime$ represents the state sent by Bob.

3) 
Then following the protocol steps, TP is required to post the message in the form of Bell states. This process can be modeled as an isometry mapping \cite{25,30} on the qubit sent by Bob and her auxiliary state, where the isometry mapping $V_2$ is defined as $V_2|j,g_{i,j}\rangle=\sum_{m=0}^{3}|m,g^{m}_{i,j}\rangle$. Thus the system can be expressed as

\begin{equation}
\begin{aligned}
\rho^{\prime}_4&=|00\rangle \langle 00|_{AB}\otimes\sum_{m=0}^3|m,g^{m}_{0,0}\rangle \langle m,g^{m}_{0,0}|_{T} \\
& +|01\rangle \langle 01|_{AB}\otimes \sum_{m=0}^3|m,g^{m}_{0,1}\rangle \langle m,g^{m}_{0,1}|_{T}\\
&+|10\rangle \langle 10|_{AB}\otimes \sum_{m=0}^3|m, g^{m}_{1,0}\rangle \langle m, g^{m}_{1,0}|_{T}\\
&+|11\rangle \langle 11|_{AB}\otimes \sum_{m=0}^3|m,g^{m}_{1,1}\rangle \langle m,g^{m}_{1,1}|_{T}.
\end{aligned}
\end{equation}
TP can observe the value of $m$ to obtain the message published to Alice and Bob, where $m=0,1,2,3$ represents that the message published by TP is $|\phi^+\rangle$, $|\phi^-\rangle$, $|\psi^+\rangle$ and $|\psi^-\rangle$, respectively.

However, Alice and Bob only accept the event where TP's measurement result is $|\phi^+\rangle$ or $|\phi^-\rangle$. That is, $m= 0,1$. Conditioning on this event, thus the final contributing raw key system is
\begin{equation}
\begin{aligned}
\rho^{\prime}_{ABT}&=\frac{1}{N^\prime}\sum_{i,j\in\{0,1\}} |i,j\rangle \langle i,j|_{AB} \otimes \sum^{1}_{m=0}|g_{i,j}^{m}\rangle \langle g_{i,j}^{m}|_{T},
\end{aligned}
\end{equation}
where $N^\prime$ is the normalization term: $N^\prime=\sum_{i,j\in\{0,1\}}\langle g^0_{i,j}| g^0_{i,j}\rangle+\langle g^1_{i,j}| g^1_{i,j}\rangle $. Then, TP may obtain the raw keys by observing the states of her auxiliary state. 

Let $p^{m}_{i,j}$ be the probability that TP sends message `$m$' and Alice and Bob's measurement results are $i$ and $j$. From (38), we have 
\begin{equation}
\begin{aligned}
&p^{0}_{0,0}=\langle g^{0}_{0,0}| g^{0}_{0,0}\rangle,&
&p^{0}_{0,1}=\langle g^{0}_{0,1}| g^{0}_{0,1}\rangle, \\
&p^{0}_{1,0}=\langle g^{0}_{1,0}| g^{0}_{1,0}\rangle,&
&p^{0}_{1,1}=\langle g^{0}_{1,1}| g^{0}_{1,1}\rangle,\\
&p^{1}_{0,0}=\langle g^{1}_{0,0}| g^{1}_{0,0}\rangle,&
&p^{1}_{0,1}=\langle g^{1}_{0,1}| g^{1}_{0,1}\rangle, \\
&p^{1}_{1,0}=\langle g^{1}_{1,0}| g^{1}_{1,0}\rangle,&
&p^{1}_{1,1}=\langle g^{1}_{1,1}| g^{1}_{1,1}\rangle.
\end{aligned}
\end{equation}
Note that the probabilities of $p^{0}_{i,i}$ and $p^{1}_{i,i}$ is higher than $p^{0}_{i,1-i}$ and $p^{1}_{i,1-i}$ ($i=0,1$); otherwise, there is too much error, and Alice and Bob will abort.
Next, let $q_{i,j}$ represent the probability that Alice's and Bob's raw keys are $i$ and $j$. It's not hard to see that
\begin{equation}
q_{i,j}=\frac{1}{N^\prime}(p^{0}_{i,j}+p^{1}_{i,j}).
\end{equation}

\subsubsection{Key rate derivation}

We have derived the final contributing raw key system $\rho^{\prime}_{ABT}$. According to the definition of key rate, as long as $r=S(A|T)-H(A|B)> 0$, a secure key can be obtained. To compute the key rate $r$, we need a bound of the entropy $S(A|T)$ and $H(A|B)$. Obviously, the latter is easy to obtain. Based to the value of $q_{i,j}$, $H(A|B)$ can be calculated as follows.
\begin{equation}
H(A|B)=H(q_{0,0},q_{0,1},q_{1,0},q_{1,1})-H(q_{0,0}+q_{1,0},q_{0,1}+q_{1,1}).
\end{equation}

Then, we need to bound the entropy $S(A|T)$. It is more challenging to calculate $S(A|T)$ in the case that TP is untrusted than in the semi-honest scenario. Unlike the approach adopted in the previous subsection, we will use the following theorem from \cite{25}:
 

\begin{theorem}
Given a quantum state $\rho^{*}_{AT}$ of the form:

\begin{equation}
\begin{aligned}
\rho^{*}_{AT}&=\frac{1}{N}\left(|0\rangle\langle 0|_{A}\otimes\left[\sum^{m}_{i=0}|E_i\rangle \langle E_i|\right] +|1\rangle\langle 1|_{A}\otimes\left[\sum^{m}_{i=0}|F_i\rangle \langle F_i|\right]\right),
\end{aligned}
\end{equation}
where $N>0$ is a normalization term. Then, it holds that:

\begin{equation}
\begin{aligned}
S(A|T)_{\rho}&\ge\sum_{i=0}^{m}\left( \frac{\langle E_i|E_i\rangle+ \langle F_i|F_i\rangle}{N} \right)\cdot 
\left(h\left(\frac{\langle E_i|E_i\rangle}{\langle E_i|E_i\rangle+ \langle F_i|F_i\rangle}  \right)-h(\lambda_i) \right)
\end{aligned}
\end{equation}
where: 

\begin{equation}
\lambda_i=\frac{1}{2}+\frac{\sqrt{(\langle E_i|E_i\rangle-\langle F_i|F_i\rangle)^2+4Re^2\langle E_i| F_i\rangle}}{2(\langle E_i|E_i\rangle+ \langle F_i|F_i\rangle)}
\end{equation}
Here, $h(x)$ means the binary Shannon entropy, i.e., $h(x) = −x log_2 x − (1 − x) log_2 (1 − x)$.
\label{thm-1}
\end{theorem}
For the detailed proof process, the reader is referred to \cite{25}. From (38), we can obtain $\rho^{\prime}_{AT}$ by tracing out $B$ from $\rho^{\prime}_{ABT}$.

\begin{equation}
\begin{aligned}
\rho^{\prime}_{AT}&=\frac{1}{N^\prime}\left(|0\rangle \langle 0|_{A} \otimes \sum^{1}_{m,j=0}|g_{0,j}^{m}\rangle \langle g_{0,j}^{m}|_{T}+ |1\rangle \langle 1|_{A} \otimes \sum^{1}_{m,j=0}|g_{1,j}^{m}\rangle \langle g_{1,j}^{m}|_{T}\right)
\end{aligned}
\end{equation}
Applying Theorem 1 to the above state, we can derive the result:
\begin{equation}
S(A|T)_{\rho}\ge\sum_{j,m}\left( \frac{\langle g^{m}_{0,j}|g^{m}_{0,j}\rangle+ \langle g^{m}_{1,1-j}|g^{m}_{1,1-j}\rangle}{N^{\prime}} \right)\cdot \left(h\left(\frac{\langle g^{m}_{0,j}|g^{m}_{0,j}\rangle}{\langle g^{m}_{0,j}|g^{m}_{0,j}\rangle+ \langle g^{m}_{1,1-j}|g^{m}_{1,1-j}\rangle}  \right)-h(\lambda_{j,m}) \right),
\end{equation}
where 
\begin{equation}
\lambda_{j,m}=\frac{1}{2}+\frac{\sqrt{(\langle g^{m}_{0,j}|g^{m}_{0,j}\rangle-\langle g^{m}_{1,1-j}|g^{m}_{1,1-j}\rangle)^2+4Re^2\langle g^{m}_{0,j}| g^{m}_{1,1-j}\rangle}}{2(\langle g^{m}_{0,j}|g^{m}_{0,j}\rangle+ \langle g^{m}_{1,1-j}|g^{m}_{1,1-j}\rangle)}.
\end{equation}
Now, we have derived the bound of $S(A|T)$, which depend on the parameters $\langle g^{m}_{0,j}|g^{m}_{0,j}\rangle$, $\langle g^{m}_{1,1-j}|g^{m}_{1,1-j}\rangle$, and $Re\langle g^{m}_{0,j}| g^{m}_{1,1-j}\rangle$. Recall (39), the values of $\langle g^{m}_{0,j}|g^{m}_{0,j}\rangle$ and $\langle g^{m}_{1,1-j}|g^{m}_{1,1-j}\rangle$ can be obtained. Thus, the next step is to bound $Re\langle g^{m}_{0,j}| g^{m}_{1,1-j}\rangle$.


Observing the case that Alice and Bob both reflect the qubit, we may bound $Re\langle g^{m}_{0,j}| g^{m}_{1,1-j}\rangle$. Specifically, 
Alice and Bob both reflect the qubit directly, so the whole transmission process can be modeled as  
\begin{equation}
\begin{aligned}
& V_{2}V_{1}\big(|0\rangle\otimes |g_0\rangle+|1\rangle \otimes|g_1\rangle \big)\\
& \quad \quad= V_{2}\big( |0\rangle \otimes|g_{0,0}\rangle+|1\rangle \otimes |g_{0,1}\rangle +|0\rangle \otimes|g_{1,0}\rangle+|1\rangle \otimes |g_{1,1}\rangle \big)\\
& \quad \quad= \sum_{m=0}^{3}|m\rangle \otimes \big(|g^{m}_{0,0}\rangle+|g^{m}_{0,1}\rangle+|g^{m}_{1,0}\rangle+|g^{m}_{1,1}\rangle \big).
\end{aligned}
\end{equation}
According the value of $m$, TP then announces the message in the form of Bell states. If the message is not $|\phi^+\rangle$, Alice and Bob know there is an error introduced. We first discuss the case that TP announces $|\phi^{-}\rangle$, i.e., $m=1$ (the case of $m=0$ is similar).

From (48), the probability that TP announces $|\phi^{-}\rangle$ can be estimated:
\begin{equation}
\begin{aligned}
p_{\phi^-}&= \big(\langle g^{1}_{0,0}|+\langle g^{1}_{0,1}|+\langle g^{1}_{1,0}|+\langle g^{1}_{1,1}| \big)  \\
& \quad \otimes \big(|g^{1}_{0,0}\rangle+|g^{1}_{0,1}\rangle+|g^{1}_{1,0}\rangle+|g^{1}_{1,1}\rangle \big).\\
\end{aligned}
\end{equation}
This provides a way to bound $Re\langle g^{1}_{0,0}| g^{1}_{1,1}\rangle$ and $Re\langle g^{1}_{0,1}| g^{1}_{1,0}\rangle$.  After some algebra, we have

\begin{equation}
\begin{aligned}
\left|p_{\phi^-}-\sum_{i.j}\langle g^{1}_{i,j}|g^{1}_{i,j}\rangle \right| &= 
\begin{vmatrix}
 2Re\langle g^{1}_{0,0}| g^{1}_{1,1}\rangle+ 2Re\langle g^{1}_{0,0}| g^{1}_{0,1}\rangle+ 2Re\langle g^{1}_{0,0}| g^{1}_{1,0}\rangle \\
 \\
+ 2Re\langle g^{1}_{1,1}| g^{1}_{0,1}\rangle+ 2Re\langle g^{1}_{1,1}| g^{1}_{1,0}\rangle+ 2Re\langle g^{1}_{0,1}| g^{1}_{1,0}\rangle
\end{vmatrix}.
\end{aligned}
\end{equation}
According to the Cauchy-Schwarz inequality, we have $Re\langle g^{1}_{0,0}| g^{1}_{1,1}\rangle \le  \sqrt{\langle g^{1}_{0,0}| g^{1}_{0,0}\rangle \langle g^{1}_{1,1}| g^{1}_{1,1}\rangle}$ and  $Re\langle g^{1}_{0,1}| g^{1}_{1,0}\rangle \le  \sqrt{\langle g^{1}_{0,1}| g^{1}_{0,1}\rangle \langle g^{1}_{1,0}| g^{1}_{1,0}\rangle}$. In general the values of $\langle g^{1}_{0,0}| g^{1}_{0,0}\rangle$ and $\langle g^{1}_{1,1}| g^{1}_{1,1}\rangle$ are higher than $\langle g^{1}_{0,1}| g^{1}_{0,1}\rangle$ and $\langle g^{1}_{1,0}| g^{1}_{1,0}\rangle$; otherwise, there is too much error, and Alice and Bob will abort. Therefore, the value of $Re\langle g^{1}_{0,0}| g^{1}_{1,1}\rangle$ is generally greater than  $Re\langle g^{1}_{0,1}| g^{1}_{1,0}\rangle$. Let $Re\langle g^{1}_{0,1}| g^{1}_{1,0}\rangle=\sqrt{\langle g^{1}_{0,1}| g^{1}_{0,1}\rangle \langle g^{1}_{1,0}| g^{1}_{1,0}\rangle}$, then we focus on bounding  $Re\langle g^{1}_{0,0}| g^{1}_{1,1}\rangle$. Following the Cauchy-Schwarz inequality and Absolute value inequality, we can obtain the range of $Re\langle g^{1}_{0,0}| g^{1}_{1,1}\rangle$

\begin{equation}
\begin{aligned}
Re\langle g^{1}_{0,0}| g^{1}_{1,1}\rangle &\ge \left|\frac{1}{2}p_{\phi^-}- \frac{1}{2}\sum_{i.j}\langle g^{1}_{i,j}|g^{1}_{i,j}\rangle \right|- \sqrt{\langle g^{1}_{0,0}| g^{1}_{0,0}\rangle \langle g^{1}_{0,1}| g^{1}_{0,1}\rangle}\\
&-\sqrt{\langle g^{1}_{0,0}| g^{1}_{0,0}\rangle \langle g^{1}_{1,0}| g^{1}_{1,0}\rangle}-\sqrt{\langle g^{1}_{1,1}| g^{1}_{1,1}\rangle \langle g^{1}_{1,0}| g^{1}_{1,0}\rangle}\\
&-\sqrt{\langle g^{1}_{0,1}| g^{1}_{0,1}\rangle \langle g^{1}_{1,1}| g^{1}_{1,1}\rangle}-\sqrt{\langle g^{1}_{0,1}| g^{1}_{0,1}\rangle \langle g^{1}_{1,0}| g^{1}_{1,0}\rangle},
 \end{aligned}
\end{equation}
where $\langle g^{1}_{i,j}| g^{1}_{i,j}\rangle$  for $i,j \in\{0,1\}$ can be estimated in (39), and $p_{\phi^-}$ can be estimated by Alice and Bob. Following this fact,  we have obtained the bounds of $Re\langle g^{1}_{0,0}| g^{1}_{1,1}\rangle$ and $Re\langle g^{1}_{0,1}| g^{1}_{1,0}\rangle$. 

For the case that TP announces $|\phi^{+}\rangle$, i.e., $m=0$, we can adopt the same method to derive the bounds of  $Re\langle g^{0}_{0,0}| g^{0}_{1,1}\rangle$ and $Re\langle g^{0}_{0,1}| g^{0}_{1,0}\rangle$. So far, we have derived the bounds of $Re\langle g^{1}_{0,0}| g^{1}_{1,1}\rangle$, $Re\langle g^{1}_{0,1}| g^{1}_{1,0}\rangle$, $Re\langle g^{0}_{0,0}| g^{0}_{1,1}\rangle$ and $Re\langle g^{0}_{0,1}| g^{0}_{1,0}\rangle$, which depend on the parameters available to Alice and Bob. Naturally, we can obtain the value of $S(A|T)$ and $H(A|B)$. Hence, the key rate under the case that TP is untrusted can be derived.

\subsubsection{Key rate evaluation }

Same as in Sect. 3.1.3, the errors introduced by TP's attacks can be characterized by the parameters $Q$, $Q_M$, and $Q_R$, and these parameters may be easily observed by users. Then, we will use these parameters to represent $\langle g^{m}_{i,j}| g^{m}_{i,j}\rangle$ and $Re\langle g^{m}_{0,j}| g^{m}_{1,1-j}\rangle$. 

Here, we use $\langle g^{1}_{0,0}| g^{1}_{0,0}\rangle$ as an example. In more detail, the probability that Alice and Bob measure the same result is $1-Q$, while the probability that TP announces $|\phi^-\rangle$ (i.e., $m=1$) is $\frac{1}{2}(1-Q_M)$ when Alice and Bob both measure. Thus, we have $\langle g^{1}_{0,0}| g^{1}_{0,0}\rangle=\frac{1}{2}(1-Q)(1-Q_M)$. It is not difficult to see (39) can be expressed as
\begin{equation}
\begin{aligned}
&p^{0}_{0,0}=\langle g^0_{0,0}|g^0_{0,0}\rangle=p^{1}_{0,0}=\langle g^1_{0,0}|g^1_{0,0}\rangle=\frac{1}{2}(1-Q)(1-Q_M),\\  
&p^{0}_{0,1}=\langle g^0_{0,1}|g^0_{0,1}\rangle=p^{1}_{0,1}=\langle g^1_{0,1}|g^1_{0,1}\rangle=\frac{1}{2}QQ_M,\\
&p^{0}_{1,0}=\langle g^0_{1,0}|g^0_{1,0}\rangle=p^{1}_{1,0}=\langle g^1_{1,0}|g^1_{1,0}\rangle=\frac{1}{2}QQ_M,\\
&p^{0}_{1,1}=\langle g^0_{1,1}|g^0_{1,1}\rangle=p^{1}_{1,1}=\langle g^1_{1,1}|g^1_{1,1}\rangle=\frac{1}{2}(1-Q)(1-Q_M).
\end{aligned}
\end{equation}
Then (40) can be rewritten as 
\begin{equation}
\begin{aligned}
&q_{0,0}=\frac{(1-Q)(1-Q_M)}{2(1-Q)(1-Q_M)+2QQ_M},&  
&q_{0,1}=\frac{QQ_M}{2(1-Q)(1-Q_M)+2QQ_M},&\\
&q_{1,0}=\frac{QQ_M}{2(1-Q)(1-Q_M)+2QQ_M},& 
&q_{1,1}=\frac{(1-Q)(1-Q_M)}{2(1-Q)(1-Q_M)+2QQ_M}. 
\end{aligned}
\end{equation}

 $Re\langle g^{m}_{0,j}| g^{m}_{1,1-j}\rangle$ for $j,m=0,1$ can also be deonted by $Q$, $Q_M$, and $Q_R$. We use $\langle g^{1}_{0,0}| g^{1}_{1,1}\rangle$ as an example for illustration, and other cases can be obtained by analogy. Applying (52) into (51), we have
\begin{equation}
\begin{aligned}
\langle g^{1}_{0,0}| g^{1}_{1,1}\rangle &\ge \left|\frac{1}{2}\big[(1-Q)(1-Q_M)+QQ_M\big]-\frac{1}{2}p_{\phi^-}\right|\\
&\quad- 2\sqrt{(1-Q)(1-Q_M)QQ_M}-\frac{1}{2}QQ_M,
 \end{aligned}
\end{equation}
where $p_{\phi^-}$ can be obtained by $Q_R$. Recall the definition of $Q_R=p_{\psi^+}+p_{\psi^-}+p_{\phi^-}$. To simplify the calculation, we assume that the probability of different errors is the same, that is, $p_{\psi^+}=p_{\psi^-}=p_{\phi^-}=\frac{1}{3}Q_R$. Hence, $r=S(A|T)-H(A|B)$ can be denoted by $Q$, $Q_M$, and $Q_R$.

To better compare with other similar protocols later,  we assume that $Q=Q_M=Q_R$. Integrating everything into the (41) and (46), the key rate $r$ can be estimated by the variable $Q$. As shown in Fig. 4, in the worst case, the noise tolerated by our protocol can reach $13.19\%$, which is better than the BB84 protocol's $11\%$ \cite{2}.

\begin{figure}[h]
\centering\includegraphics[width=4in]{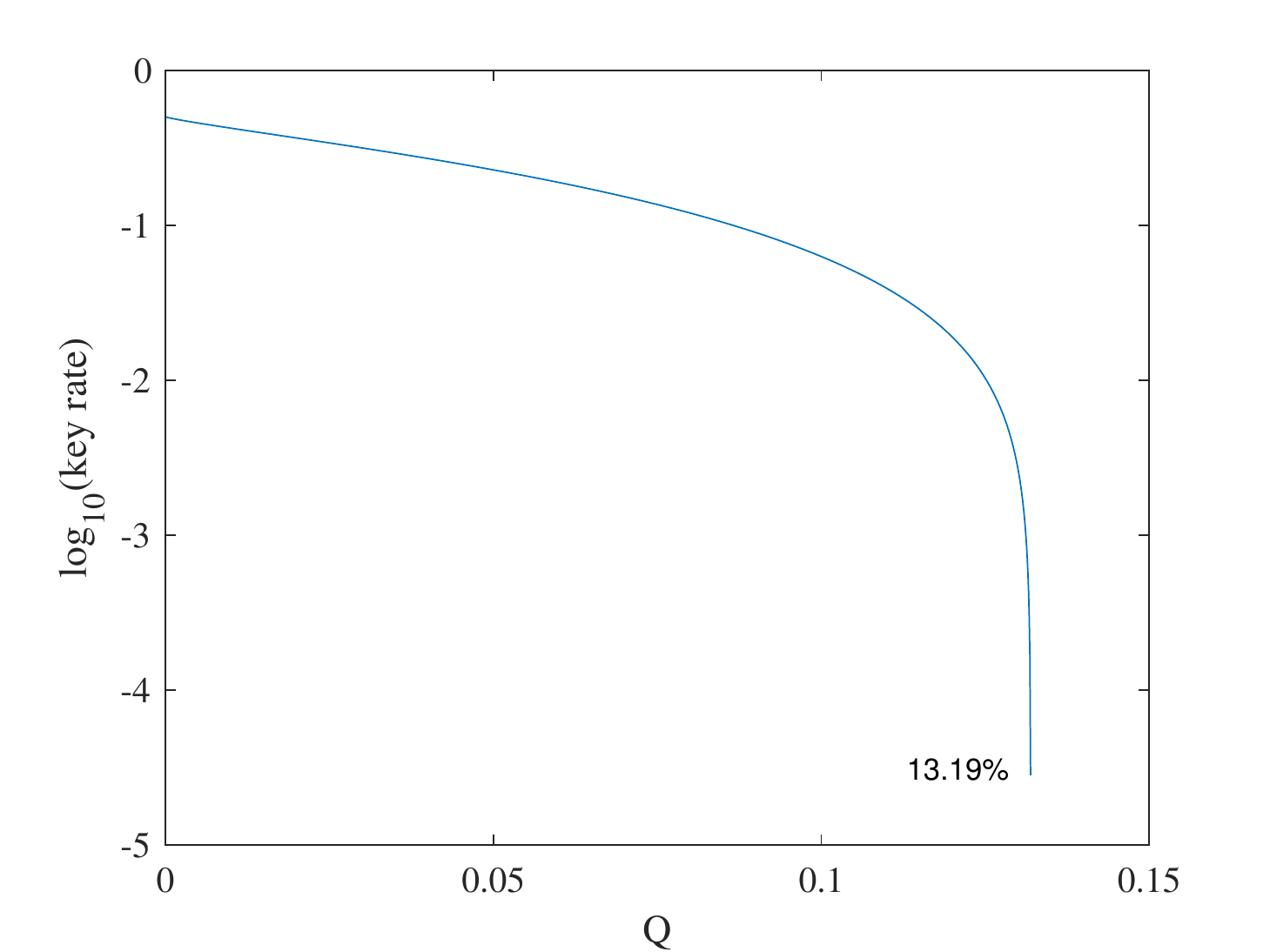}
\caption{ A diagram of our protocol's key rate when TP is untrusted. Observe that the protocol can work with the noise of less than 13.19\%. }
\label{fig:2}       
\end{figure}

\section{Performance analysis and Comparison }
\label{sec:4}
Here, we evaluate the performance of our protocol in terms of noise tolerance, qubit efficiency, communication cost, and scalability.

As an essential indicator of the semi-quantum protocol, noise tolerance is the focus of our attention. In the above analysis, we have given the noise tolerance of the protocol under different conditions, respectively. In particular, if TP is semi-honest (i.e., TP follows the protocol description thoroughly, but otherwise, she can do anything she likes), our protocol's noise tolerance is $16.02\%$. If TP is untrusted (TP can do anything she wants, including preparing fake particles and publishing fake results), our protocol's noise tolerance can reach $13.19\%$, which is better than the BB84 protocol's $11\%$. The results show that our M-SQKD protocol has similar security as the full quantum protocol.

Qubit efficiency is defined as the ratio of the number of raw keys to the total number of qubits consumed \cite{28}. The previous protocols mainly considered the noiseless case when calculating qubit efficiency\cite{24,25,26,27,28,29,30,31,32}. To better highlight the advantages and disadvantages of qubit efficiency between different protocols, this paper will calculate the qubit efficiency under noiseless conditions. In this paper, generating $1$-bit raw key, TP should prepare $4$ Bell states in the form of $|\phi^+\rangle$, and two ``classical'' users, Alice and Bob, need to prepare $2$ qubits to substitute their measured qubits, respectively. Following this, our protocol's qubit efficiency is $\frac{1}{12}$. 


Communication cost is also a metric to evaluate protocol performance. It refers to the number of qubits that need to be transmitted to generate one bit of the raw key\cite{13}. The transmission of qubits in our protocol is from TP to Alice, to Bob, and then back to TP. Thus, for producing a 1-bit raw key, four qubits (TP should prepare $4$ Bell states and send the second qubit of these Bell states to Alice ) need to be transmitted three times between TP, Alice, and Bob. That is, 12 qubits are transmitted for the whole process. 

Lastly, we consider scalability as one of the protocol performance metrics. If a protocol has good scalability, there is no doubt that it will be applicable to more scopes and scenarios. Due to the characteristics of the circular transport structure, our protocol is well scalable and can be easily extended to more than two ``classical'' users' key distribution scenarios (which will be described in the next subsection). 

For clarity, we compare similar M-SQKD protocols \cite{24,25,26,27,28,30} from the above aspects. A detailed comparison is shown in Table 1. It's clear from the table that our protocol's noise tolerance is close to previous protocols, and our protocol can be applied to more than two ``classical'' user scenarios, while previous protocols are only applicable to two ``classical'' user scenarios. In addition, our protocol still has some advantages in terms of qubit efficiency and communication cost.


\begin{table}[!htp]
\centering 
\tabcolsep 10pt
\caption{Comparisons among similar M-SQKD protocols }
\begin{threeparttable}
\label{tab:1}       
\begin{tabular}{ccccc}
\hline
              &  Noise tolerance under & Qubit & Communication & Scalability \\
              &  untrusted TP   &  efficiency &  cost &\\              
  \hline
  \noalign{\smallskip}
      Ref. \cite{24}  &  $10.65\%$  & $\frac{1}{24}$ & $32$-qubits  & \ding{53} \\
      \noalign{\smallskip}
      Ref. \cite{25}  &  $13.04\%$  & $\frac{1}{24}$ & $32$-qubits  & \ding{53}\\
      \noalign{\smallskip}
      Ref. \cite{26}  &  /  & $\frac{1}{16}$ & $24$-qubits  & \ding{53}\\
      \noalign{\smallskip}
      Ref. \cite{27}  &  /  & $\frac{1}{24}$ & $32$-qubits  & \ding{53}\\
      \noalign{\smallskip}
      Ref. \cite{28}  &  /  & $\frac{1}{8}$ & $12$-qubits  & \checkmark\\
      \noalign{\smallskip}
      Ref. \cite{30}  &  $ 9.1\%$  & $\frac{1}{12}$  & $16$-qubits  & \ding{53}\\
      \noalign{\smallskip}
      Our protocol   &  $13.19\%$  & $\frac{1}{12}$  & $12$-qubits  &  \checkmark\\
\hline
\noalign{\smallskip}
\end{tabular}
\begin{tablenotes}
\footnotesize
\item Note that Refs. \cite{26,27,28} only give security analysis under several specific attacks and no noise tolerance is calculated for such protocols. 
\end{tablenotes}
\end{threeparttable}
\end{table}

\section{Extention of the protocol to multi-party scenarios }
  
 Our protocol adopts the circular transport structure. The advantage of this structure is that it makes the protocol very flexible. Following this fact, our protocol can be easily expanded to multi-party scenarios, allowing the key distribution between more than two “classical” users.
 Therefore, it can be successfully used for multi-party M-SQKD and multi-party SQKD.

\subsection{The multi-party M-SQKD protocol}
\label{sec:5}
Our protocol can be developed into a multi-party M-SQKD protocol involving $L (L>2)$ ``classical'' users. We use $C_1,C_2,\cdots,C_L$ to denote these  ``classical'' users. Only TP is a fully quantum user in this protocol, while other users are both ``classical'' whose quantum power is limited. The specific steps are as follows.

\textbf{Step 1:} TP prepares $2^{L+1}(N+\delta)$ Bell states in the form of $|\phi^+\rangle$, where $\delta$ is a fixed parameter. Then she sends the second qubit of each Bell state to $C_1$.	
		 			
\textbf {Step 2:} For $l=1,2,\dots,L-1:$ $C_l$ randomly decides to reflect the qubit to $C_{l+1}$ or measure and resend the qubit to $C_{l+1}$.

\textbf {Step 3:} Upon receiving qubit from $C_{L-1}$, $C_{L}$ randomly chooses to reflect it to TP or measure and resend it to TP.

\textbf {Step 4:} TP adopts the Bell basis to measure the received qubits and those left in her hands. Then, TP announces her measurement results to $C_1,C_2,\cdots,C_L$.

\textbf {Step 5:} After TP releases her measurements, $C_1,C_2,\cdots,C_L$ publish their choices through the authenticated classical channel. 
\begin{itemize}
\item [\textbf{a)}] If all $C_1,C_2,\cdots,C_L$ reflect the received qubit, TP should always get $|\phi^+\rangle$. This case is used for checking TP's honesty and eavesdropping. When the error rate of this case surpasses the threshold, the protocol ends.

\item [\textbf{b)}] If all $C_1,C_2,\cdots,C_L$  measure and resend the received qubit, $C_1,C_2,\cdots,C_L$ will use these measurement results as their raw keys when TP's announcement is $|\phi^+\rangle$ or $|\phi^-\rangle$.

\item [\textbf{c)}] If their operations are inconsistent, $C_1,C_2,\cdots,C_L$ will discard these qubits.
\end{itemize}

\textbf {Step 6:} $C_1,C_2,\cdots,C_L$ randomly pick out a small subset to publish their measurement results for eavesdropping detection. Once the error rate is over the threshold, the protocol ends. If not, the protocol continues.

\textbf {Step 7:} After the classical post-processing, the ``classical'' users $C_1,C_2,\cdots,C_L$ can obtain the final secure key.

\subsection{The multi-party SQKD protocol}
The previous subsection gave detailed steps for the multi-party M-SQKD protocol. In this part, we can extend the protocol to a multi-party SQKD protocol by slightly changing the steps. Note that in the SQKD protocol, the quantum user is fully trusted, so here we redefine TP, which is trusted in this protocol. The revised steps are detailed below, while others are the same as described in Sect. 5.1.

\textbf{Step 4*}  After confirming that TP has obtained the qubits,  $C_1,C_2,\cdots,C_L$ publish their choices through the authenticated classical channel.

\textbf{Step 5*}  TP performs different operations according to the following cases:
\begin{itemize}
\item  (i) If all $C_1,C_2,\cdots,C_L$ reflect the received qubit, TP will measure the received qubit and the qubit in her hand with the Bell basis. This case is used for checking the eavesdropping. Once the error rate of this case surpasses the threshold, the protocol ends.
 
 \item (ii) If all $C_1,C_2,\cdots,C_L$ measure and resend the received qubit, TP will measure the received qubit with $Z$ basis.  This case will be used for generating the raw keys.
 
 \item (iii) If $C_1,C_2,\cdots,C_L$ choose different operations, this case will be discarded.
 
\end{itemize}

\textbf{Step 7*}  After the classical post-processing, TP and $C_1,C_2,\cdots,C_L$ can obtain the final secret key.

In this section, we extend the protocol to multi-party M-SQKD and multi-party SQKD protocols. The results show that the protocol has good scalability to enable key distribution among multiple classical users. 

\section{Conclusion }
\label{sec:6}
In this paper, we design a scalable mediated semi-quantum key distribution protocol, which allows more than two ``classical'' users to establish a secret key with the help of an adversarial TP. Then, we give a detailed security proof to show the security of this circular M-SQKD protocol. By utilizing the parameters observed in the channel, a lower bound of the protocol's key rate and noise tolerance can be derived. The classical users can distill a secure key when the noise is less than the threshold value. Indeed, even if TP is untrusted, our protocol's noise tolerance is better than the BB84 protocol. That is, our protocol holds similar security to a fully quantum one. Compared with previous M-SQKD protocols, our protocol has comparable noise tolerance and advantages in qubit efficiency and communication cost. In addition, our protocol can work on more than two ``classical'' user scenarios, while previous protocols are only applicable to two ``classical'' user scenarios. Due to these features, this protocol may have potential applications in future quantum communications.
\\
\\
\\
\textbf{Acknowledgments} 
This work was supported by the Open Research Fund of Key Laboratory of Cryptography of Zhejiang Province No. ZCL21006, and the BUPT Excellent Ph.D. Students Foundation No. CX2021117.

\end{document}